\documentclass[journal=jacsat,manuscript=article]{achemso}
\usepackage[version=3]{mhchem} 
\usepackage{xcolor}
\usepackage{multirow}

\author{Anand Dev Ranjan}

\affiliation[IISER]
{Department of Physical Sciences, Indian Institute of Science Education and Research, Kolkata, Mohanpur, 741246, India}

\author{Rakesh Sen}
\affiliation[IISER]
{EFAML, Materials Science Center, Department of Chemical Sciences, Indian Institute of Science Education and Research, Kolkata, Mohanpur, 741246,  India}

\author{Rahul Vaippully}
\affiliation[IIT M]
{Department of Physics, Indian Institute of Technology Madras, Chennai, 600036, India}

\author{Sumeet Kumar}
\affiliation[IISER]
{Department of Physical Sciences, Indian Institute of Science Education and Research, Kolkata, Mohanpur, 741246, India}

\author{Soumya Dutta}
\affiliation[IIT M]
{Department of Electrical Engineering, Indian Institute of Technology Madras, Chennai, 600036, India}

\author{Basudev Roy}
\email{basudev@iitm.ac.in}
\affiliation[Unknown University]
{Department of Physics, Indian Institute of Technology Madras, Chennai, 600036, India}

\author{Goutam Dev Mukherjee}
\affiliation[IISER]
{National Center for High Pressure Physics, and Department of Physical Sciences, Indian Institute of Science Education and Research, Kolkata, Mohanpur, 741246, India}

\author{Soumyajit Roy}
\email{s.roy@iiserkol.ac.in}
\affiliation[IISER]
{EFAML, Materials Science Center, Department of Chemical Sciences, Indian Institute of Science Education and Research, Kolkata, Mohanpur, 741246, India}

\author{Ayan Banerjee}
\email{ayan@iiserkol.ac.in}
\affiliation[IISER]
{Department of Physical Sciences, Indian Institute of Science Education and Research, Kolkata, Mohanpur, 741246, India}

\title[An \textsf{achemso} demo]
  {Giant conductance of PSS:PEDOT micro-surfaces induced by microbubble lithography}

\abbreviations{IR,NMR,UV}

\begin{document}

\begin{abstract}
  We provide direct evidence of the effects of interface engineering  of various substrates by Microbubble lithography (MBL). We choose a model organic plastic (or polymer) poly(3,4-ethylenedioxythiophene) polystyrene sulfonate (PEDOT:PSS), with conductivity of 140 S/cm, as a representative organic system to showcase our technique. Thus, we fabricate permanent patterns of PEDOT:PSS on glass, followed by a flexible PDMS substrate, and observe  conductivity enhancement of 5 times on the former (694 S/cm), and 20 times (2844 S/cm) on the latter, without the use of external doping agents or invasive chemical treatment. Probing the patterned interface, we observe that MBL is able to tune the conformational states of PEDOT:PSS from coils in the pristine form, to extended coils on glass, and almost linear structures in PDMS due to its more malleable liquid-like interface. This results in higher ordering and vanishing grain boundaries leading to the highest conductivity of PEDOT:PSS on PDMS substrates. 
\end{abstract}

\section{Introduction}
Light-matter interactions at mesoscopic length scales have led to fascinating  emergent phenomena that have modified the properties of both light and matter, and have therefore led to deep physical understanding of nature itself\cite{tsesses2019nanoletters,hertzog2019chemicalsocreviews,giannini2010small}. Thus, spin-orbit interactions of light  have been driven by plasmonic materials\cite{tsesses2019nanoletters}, while  the chemical and physical properties of molecules have been modified by coupling them strongly with light\cite{hertzog2019chemicalsocreviews,tiwari2020JMCC}. Besides fundamental studies, multifarious applications have also been facilitated by light-matter interactions, ranging from communications and quantum information processing\cite{shao2018natcom}, to exciting applications in nano-optics and optoelectronics\cite{mueller2018npj2dmaterials}. Even at somewhat larger length scales, the interaction of light with nanoscopic matter proves to be very useful for applications in  biological light harvesting systems, and for the development of efficient artificial photovoltaic devices\cite{lienau2014Jopt}. In this context, interfaces play a crucial role. Interfacial structure and emergent properties thereof can be induced by simple experimental heuristics rendering such techniques indispensable\cite{zhou2017naturenanotech,lee2010chemmaterials,kim2021angewandte,brus2016organicelecronics}.

In the case of plastic electronics - a crucial factor is the conductivity of the material used. While conductive organic polymers are typically employed for this purpose, their conductance is often limited by their electronic structures\cite{walton1990matdesign, nezakati2018chemreviews}. Efforts to improve the conductivity of the polymers extrinsically typically rely on doping\cite{ma2019chemscience,ahmad2021materialsadvances} and chemical treatment\cite{ponder2022JMCC}. In addition, all these methods consist of multiple processing stages and are therefore time consuming, while doping with the help of harsh chemicals also renders the material entirely unsuitable for any biological application. With regard to this, the role of interfaces in affecting the conductivity of such polymers remains an open question. With various tools of interface engineering being available, such dependence could prove to be of considerable significance in enhancing conductivities of conductive polymers in a generic and non-invasive manner. 

Microbubble lithography (MBL) has established itself as a robust tool in self-assembly based bottom-up approaches in fabricating micro-patterns on transparent substrates for use in diverse applications including plastic electronics, catalysis, and even biodetection\cite{ghosh2020nanoletters,armon2021wiley,thomas2015site,zheng2020nanoletters}. Especially in the context of plastic electronics, MBL has been used to pattern conductive polymers and even increase their conductivity significantly by doping occurring concurrently with the patterning process\cite{ghosh2017JMCC}. On another note, it has been demonstrated that irreversible self assembly under non-equilibrium conditions is a reliable method for carrying out interface engineering\cite{mann2009nature}. MBL promises to be an excellent candidate for implementing such designs. Indeed, several observations in MBL seem to suggest fundamental changes in the properties of mesoscopic matter post patterning - thus leading to exciting studies to reveal the intrinsic science behind the observations\cite{li2021ACS,ghosh2020nanoletters}. Thus, it appears worthwhile to investigate the role of interfaces in the conductivity of conducting polymers patterned by MBL. This is because MBL makes it possible to restructure interfaces, which allows their chemical modifications to appear as emergent phenomena. Understandably, the presence of different surface matrices would further influence and enhance these phenomena by offering a modular approach for influencing the intrinsic properties of matter, of which one is conductivity.

In this paper, we investigate the role of substrates in affecting the conductivity of micropatterned organic conducting polymers using MBL.
We selected the organic polymer PEDOT: PSS (poly(3,4-ethylenedioxythiophene)) as a representative model system since it has been extensively investigated, and there are several reported results of its application as diodes, PLEDs, supercapacitors, and numerous other devices\cite{da2013IEEE,cook2014JMCC,manjakkal2020Wiley,sun2015Springer}. Indeed, the material has been identified as one of the primary candidates for the development of next generation flexible plastic electronics\cite{wen2017wiley}. However, PEDOT-PSS also has low conductivity in its pristine form akin to most organic conducting polymers as mentioned earlier,  which renders it virtually unsuitable for any electronic device fabrication without extrinsic strategies to improve conductivity. For PEDOT-PSS, besides the usual approaches including doping and chemical treatment, the conductivity has also been improved by applying mechanical stress, blending with other NPs, etc.\cite{sarkar2017softmatter,shi2015Wiley}. However, in all other methods - to the best of our knowledge, the increase in conductivity is performed independently of the patterning strategy, and are two different processes. On the other hand, MBL - by character, juxtaposes the patterning and conductivity-increase processes\cite{ghosh2017JMCC} - which is a significant advantage in process flow optimization for actual applications. Thus, we pattern PEDOT-PSS in two different transparent substrates - glass and PDMS - and observe significant conductivity increase in both cases over the pristine sample. We attempt to understand the underlying physical reason behind such increase of conductivity with the help of various characterization tools including scanning electron microscopy coupled with energy dispersive Xray spectroscopy (SEM-EDX), Raman spectroscopy, and cyclic voltametry. A careful investigation of the results from all the measurements point to conformational and morphological changes in the PEDOT-PSS crystal structure itself due to high absorption at the laser wavelength used to drive MBL. Indeed,  the large energy absorbed due to the high light intensities generated in MBL, melts the insulating PSS shells encapsulating the PEDOT cores. This increases intercrystalline boundaries and grain size of the polymer as we pattern from glass to PDMS - so that the diffusion lengths of charge carries increase as we change the substrate, resulting in higher conductivities being observed. Our results show that the choice of substrates are an important tool to improve the conductivity of organic polymers, and MBL - with its innate capability to simultaneously pattern and induce morphological and conformational changes of the material being patterned so as to increase conductivity - can be an ideal candidate to facilitate high-performing plastic electronics. 

\begin{figure}[!h]
\centering
\includegraphics[width=16cm]{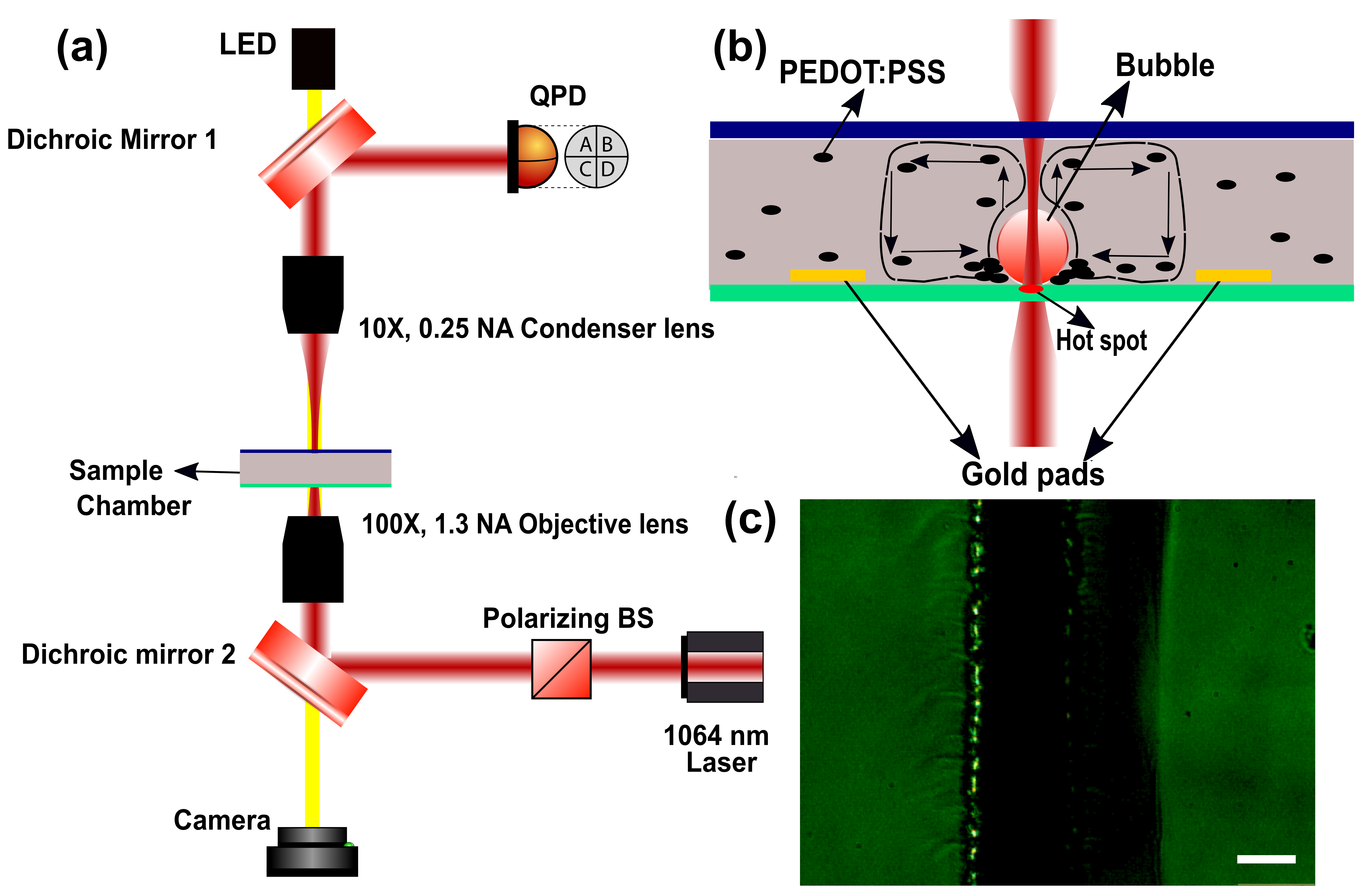}
\caption{Experimental Setup: (a) Schematic of experimental setup used for the experiment. A 1064 nm laser beam is focussed on the sample chamber using 100x (1.3 NA) objective lens and a set of dichroic mirror (Dichroic Mirror 2) and polarising beam splitter (Polarising BS). The sample chamber is imaged using a visible light source (LED) which falls on the sample chamber and is collected by the camera (b) When the focused laser beam is incident on the adsorbed material on glass slide it creates a hot spot resulting in formation of microbubble. The microbubble induces a Gibbs-Marangoni flow in the dispersed material as shown by arrows. This results in deposition of material at the base of the microbubble (c) The patterned PEDOT:PSS on glass substrate using the MBL.}
\label{schematic}
\end{figure}

\begin{figure}[!h]
\centering
\includegraphics[width=15cm, height=12cm]{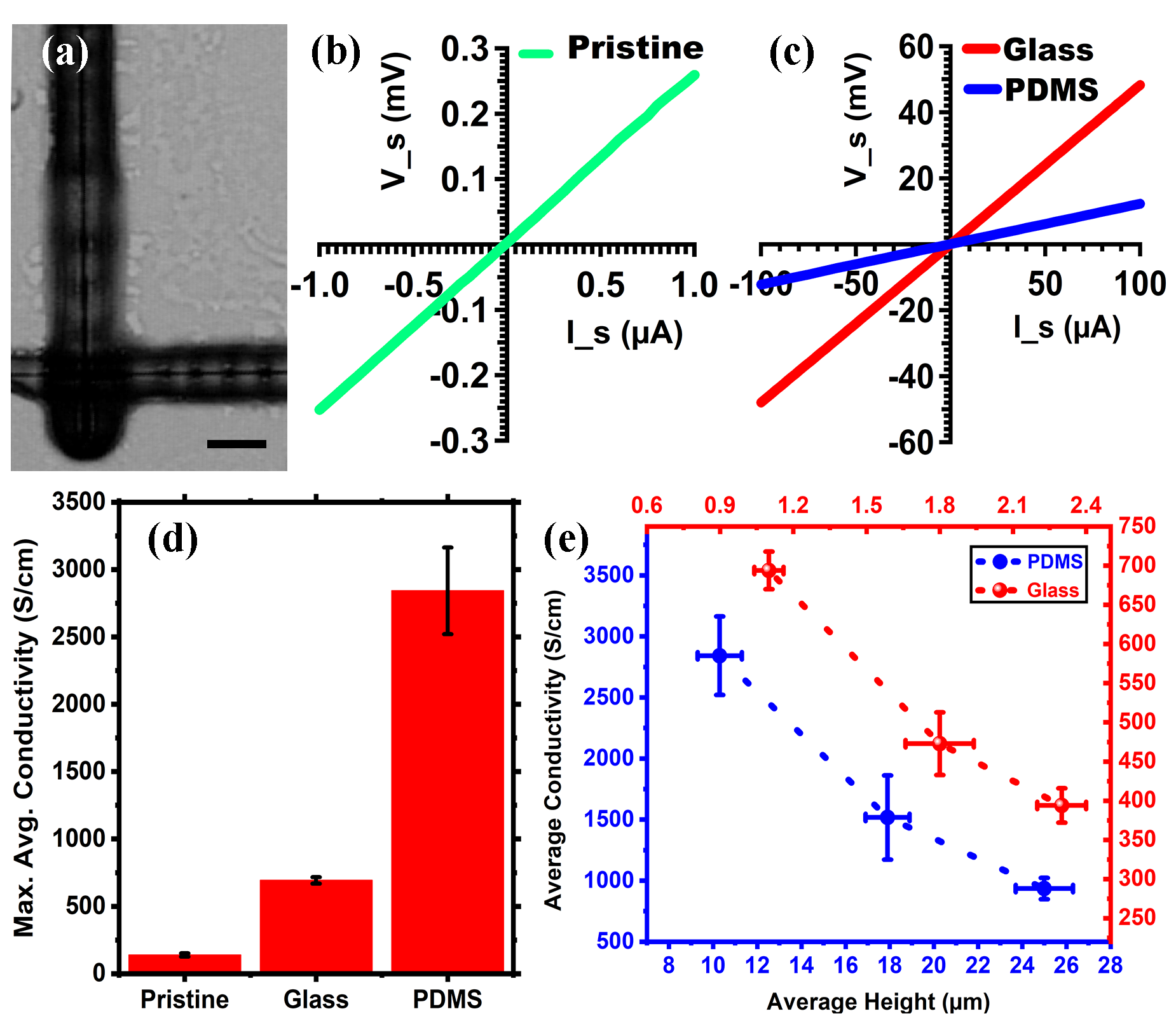}
\caption{(a) Brightfield image of patterned PEDOT:PSS in a Hall-bar geometry (b) Conductivities of pristine and patterned PEDOT:PSS on glass and PDMS. Patterns on the PDMS substrate show the maximum conductivity compared to either of the samples. The conductivity of pristine is lowest of all the samples (c) IV characteristics of pristine PEDOT:PSS (d) IV graph of the patterned PEDOT:PSS on glass and PDMS substrate.}
\label{glassiv}
\end{figure}

{\it Results and Discussions: I. Experimental setups for patterning and subsequent measurements on the patterns:} The schematic of the setup used for the experiments is shown in Fig.~\ref{schematic}a. The patterning was performed using an Olympus 1X71 microscope equipped with an inverted Plan-Fluorite 100x oil immersion objective with 1.3 Numerical Aperture (NA) and an overfilled illuminating aperture. The 1064 nm diode laser (Lasever) is  focused at the sample plane by the 100x objective, and illumination is carried out using white light passing through a condenser lens, which is separated from the back-scattered IR laser using dichroic mirrors placed before the camera that is fixed at the back-focal plane of the objective lens. The sample chamber, as shown in Fig.~\ref{schematic}b is filled with 10 $\mu$l of PEDOT:PSS obtained from Sigma Aldrich. The detailed methodology behind MBL has been already described  earlier\cite{roy2013langmuir,ghosh2017JMCC,ghosh2020nanoletters}. Here, we use different laser powers (ranging from 3-30 mW at the focal plane) to grow microbubbles of different sizes, and translate them by moving the microscope sample-holder stage, so as to obtain PEDOT:PSS patterns of varying widths both on glass and PDMS substrates. Fig.~\ref{schematic}c demonstrates a linear pattern on glass, while Fig.~\ref{glassiv}a shows a Hall-bar structure patterned on PDMS. Both have been drawn using a laser power of around 24 mW. 

After patterning, we performed a set of experiments to measure the conductivity of the patterns, and provide explanations for the enhanced conductivity we observed. Thus, we measured the conductivity of the samples at room temperature by the four-point probe technique using a Keithley 2400 Source Meter Unit. Further, scanning electron microscope (SEM) and Energy Dispersive X-Ray Spectroscopy (EDX) of the patterned material were performed using the Zeiss GeminiSEM using $K_\alpha$ lines of Cu. We then performed Raman measurements using a Horiba micro-Raman spectroscopy system in the back scattering configuration. The Raman lines were excited using a 532.2 nm laser with an incident power of 1 mW and resolution around 1.2 cm-1. Finally, cyclic voltammetry was performed with a workstation (CH Instruments, Model CHI7091E) in a three-electrode electrochemical setup where 0.1 M sodium sulphate solution is used as an electrolyte. The fabricated patterns are used as the working electrode along with platinum wire as a counter electrode and Ag/AgCl (1M KCl) as reference electrode. 

{\it II. Conductivity measurements:} To determine the conductivity of the PEDOT:PSS patterns deposited on glass and PDMS, we measured the current-voltage characteristics as shown in Fig.~\ref{glassiv}c-d \cite{ghosh2017JMCC}, and employed a four-probe measurement, where voltage was provided to one arm of the pattern, while the current was measured in the perpendicular arm as shown in Fig.~\ref{glassiv}a. Conductivities of all the samples together is shown in Fig.~\ref{glassiv}b with the corresponding highest and lowest values. We found that the pristine sample has average conductivity of around $140\pm 12 $ S/cm (mean $\pm~ 1\sigma$) with the lowest conductivity of roughly 141 S/cm, whereas patterns on glass had a 5 times increase in conductivity, with an average conductivity of around $694\pm 24$ S/cm. The value of the conductivity is already almost three times higher than what we had achieved with  other well-known conducting polymers - polypyrrole and polyaniline - that were similarly micropatterned on glass \cite{ghosh2017JMCC} - but in conjunction with soft oxometalates. We  observed an even more enhanced conductivity of the PEDOT:PSS micropatterns that were fabricated on PDMS substrates - the average conductivity being $2844\pm 321$ S/cm. Note that this is comparable to the highest conductivity of PEDOT:PSS reported in the literature after patterning on a PDMS substrate (around 3605 S/cm) - this being achieved by doping with sulphuric acid\cite{shi2015Wiley}. Our process, obviously, does not involve any such chemical treatment, and is clearly a consequence of the patterning process itself. The question, however, arises as to why the same material when patterned on multiple substrates exhibits different values of conductivity. This is what we attempt to answer now,  using different experimental probes, and  look for any physical or chemical changes in PEDOT:PSS that could cause such  significant change in an intrinsic quantity such as conductivity.

\begin{figure}[!h]
\centering
\includegraphics[width=16cm]{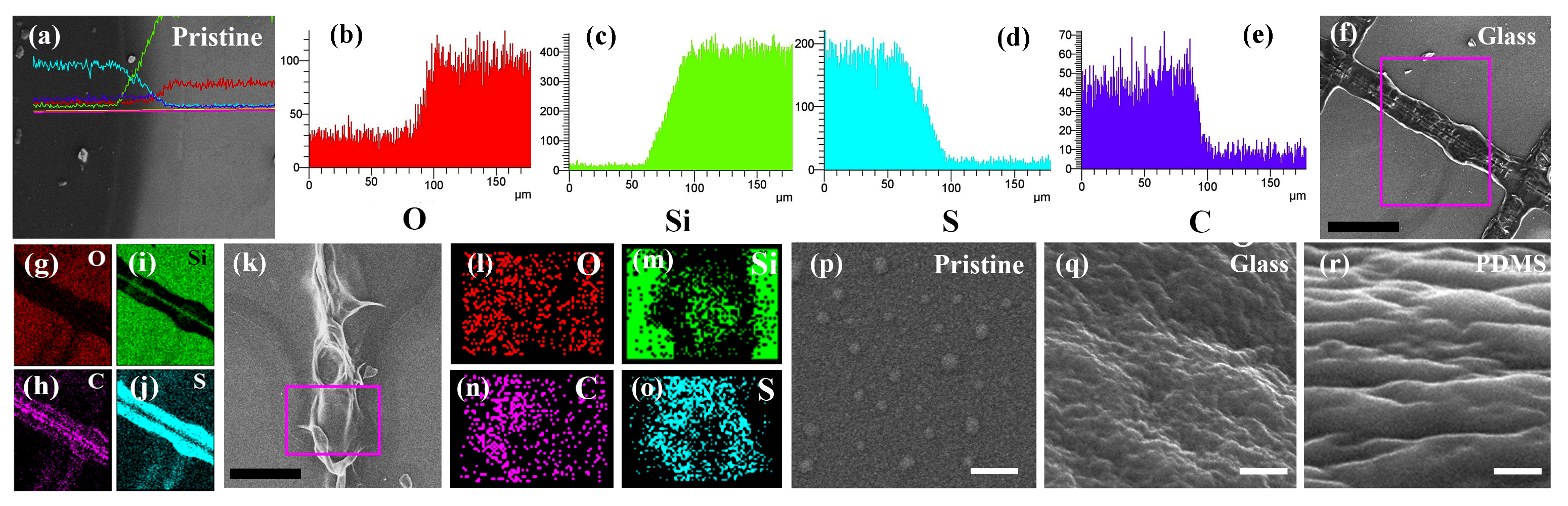}
\caption{To determine the elemental composition along the patterned region of the PEDOT:PSS we used the SEM-EDX. This image maps the elemental constituents of PEDOT:PSS i.e Oxygen(O), Silicon(Si), Carbon(C) and Sulphur(S), if present at the given mapped region. (a) We take the elemental map of the pristine PEDOT:PSS, where (b) to (e) shows the presence of O, Si, S and C respectively in the pristine sample. (f) PEDOT:PSS pattern on glass substrate. (g) to (j) show EDX mapping of the pattern on glass showing the presence of O, C, Si and S, respectively, along the patterned regions. (k) PEDOT:PSS pattern on PDMS substrate. (l) to (o) display elemental mapping of O, C, Si and S, respectively, present in the patterned area. We conclude from this image that the patterned region shows high concentration of the elemental constituents of PEDOT:PSS in both the substrates (p-r) the surface morphology of PEDOT:PSS present in the three configurations (scale bar represents 300 nm) (p) The SEM picture of the pristine sample clearly shows the small grain size with non-continuous boundaries which increases for the case of patterns on glass substrate as shown in (q) The PDMS patterns (r) shows longer and larger grain size compared to all other samples with few non-continuous domains.}
\label{SEM-Morpho}
\end{figure}

 To determine the underlying process behind the increase of conductivity in our case, we employed various experimental techniques such as scanning electron microscopy (SEM), EDX, Raman spectroscopy and cyclic voltametry (CV) to examine the pristine (taken in the form of a thin film on a glass cover slip) sample, and those patterned using MBL on glass, and PDMS substrates, respectively. The SEM/EDX images of the pristine and the patterned PEDOT:PSS on glass and PDMS are shown in Fig.~\ref{SEM-Morpho}a-o, with the SEM images displayed in Fig.~\ref{SEM-Morpho}a, f $\&$ k, respectively, while the EDX elemental mapping of the pristine sample, and the patterns on glass and PDMS are shown in Fig.~\ref{SEM-Morpho}b-e, g-j $\&$ l-o, respectively. These images proves the continuity and are also consistent with the expected chemical composition of elements in a PEDOT:PSS sample comprising of Oxygen (O), Carbon(C), Silicon (Si) and Sulphur (S)\cite{greczynski1999Elsevier}. 

{\it Investigations to explain high conductivity - analysis of SEM/EDX data:} The conductivity of PEDOT:PSS is also dependent on the inter-crystalline boundaries and grain size, as is well-known in the literature\cite{huang2003Elsevier}. For our samples, we used the SEM data  to look for any changes in grain size and morphology. We can clearly see from Fig.~\ref{SEM-Morpho} that the domain size of PEDOT:PSS has increased from pristine (Fig.~\ref{SEM-Morpho}p) to patterned samples (Fig.~\ref{SEM-Morpho}q-r). This increase in size of domains can be attributed to the conformational change in the PEDOT:PSS as we go from pristine to the patterns on glass and PDMS. In addition, it is clear from the SEM images (Fig.~\ref{SEM-Morpho}q) that in comparison to the other two samples, the pristine sample has exceptionally sharp grain boundaries. There are also regions in the pristine thin film where the inter grain spacing is higher (black lines). When compared to the pristine sample, the glass samples have fewer defined grain boundaries, and grains are larger, and virtually continuous. In the case of the PDMS sample, grain size increases even further and grain boundaries practically disappear. Now, grain boundaries act as solid walls, preventing charges from diffusing over the patterns and thin layer. As a result, the  reduction in the  number of grain boundaries and increase in the grain size improves charge carrier diffusion across the PEDOT:PSS patterns on glass substrates compared to the pristine sample, and even more on PDMS substrates\cite{spivak2022improving, huang2021effect} compared to the others.

\begin{figure}[!h]
\centering
\includegraphics[width=17cm]{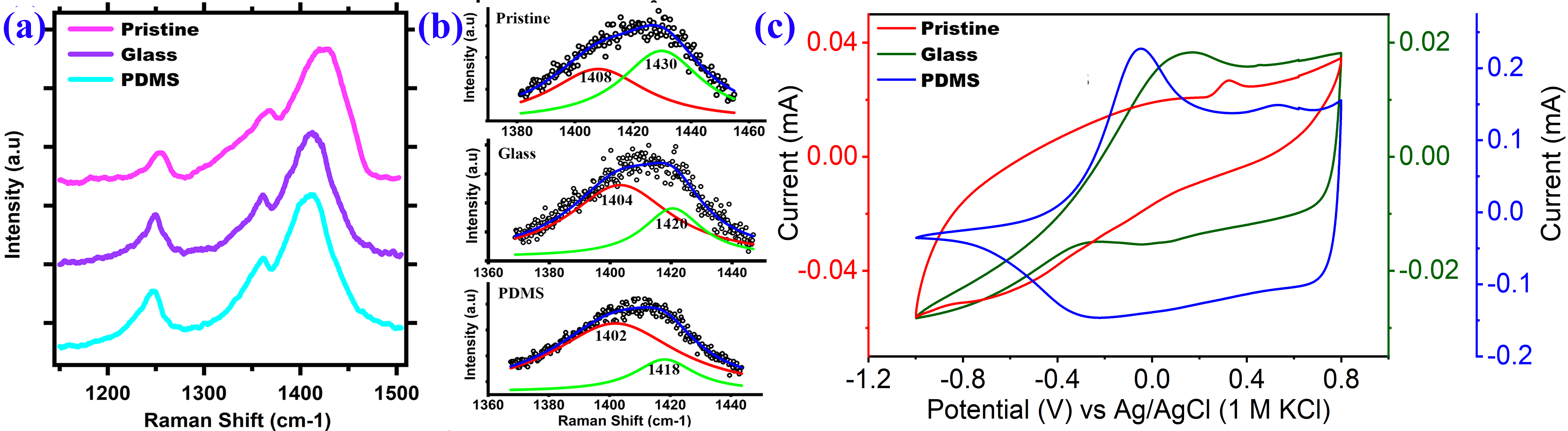}
\caption{(a) Raman spectra of pristine PEDOT:PSS and patterns on glass and PDMS (b) deconvolution of the broader peak around 1420 into two Lorentzian peaks where the smaller wave number peak correspond to the quinoid mode and the higher wavenumber peak denotes the benzoid modes of vibrations of the PEDOT:PSS samples (c) cyclic voltammogram of pristine PEDOT:PSS and the pattern on glass and PDMS recorded against Ag/AgCl (1 M KCl) at a scan rate of 50 mV/s.}
\label{Raman-cv}
\end{figure}

{\it Investigations to explain high conductivity by Raman spectroscopy:}  Although the EDX data validates the elemental composition and presence of PEDOT:PSS inside the patterned region, they do not provide evidence for conformational and molecular changes. On a similar note, SEM pictures reveal the morphology of the surface -  validating the hypothesis that patterning on different substrates leads to surface alterations - but they cannot display volume changes. To glean information regarding these aspects, as well as to corroborate doping in the case of PDMS, we performed Raman spectroscopy on the pristine and patterned samples on the two different substrates, as shown in Fig.~(\ref{Raman-cv}). Earlier studies have pointed out that the Raman fingerprints between 1200 to 1500 is of considerable importance to judge the conformation and structural property of the PEDOT:PSS chains\cite{}. The Raman vibrations around 1250 cm$^{-1}$ correspond to the $C_a$-$C_a$ inter-ring stretching, while the broader mode around 1425 cm$^{-1}$ is identified as the $C_a$=$C_b$ symmetric stretching modes of the five-member thiophene ring of the PEDOT:PSS \cite{kim2016Elsevier}. In our case, these two peaks can be clearly seen in all spectra - however, their characteristics appear to change significantly for the patterned PEDOT:PSS in comparison to the pristine sample.  

We focus first on the inter-ring stretching mode at 1250 cm$^{-1}$. The spectra clearly reveal two changes: the first is a gradual rise in the relative intensity profile by $\sim 40\%$, and the second is a red shift of the peaks from the pristine (1244 cm$^{-1}$ ) to the PDMS (1252 cm$^{-1}$ ) samples, as seen in Fig.~\ref{Raman-cv}a and Table~\ref{raman-1250}. The same characteristics are observed when PEDOT:PSS is annealed at a temperature between 200 and 300 $^{0}$C, which denotes the melting of the PSS surrounding PEDOT, resulting in improved contact between PEDOT chains\cite{schaarschmidt2009JPCB}. Importantly, similar or higher temperatures are attained while nucleating a microbubble\cite{ghoshnjc} - which implies that MBL performs a role akin to annealing here. Now, this increase in the normalised integrated intensity of the Raman modes has been linked to the subsequent lengthening of the PEDOT chain, which is related to the decrease in material resistivity shown in the electrical characterisation \cite{rutledge2015ACSAMI}. This indicates that the microbubble causes melting of the PSS structures, thus increasing the contact between the conductive PEDOT chains and providing a boost to the charge conduction mechanism.

\begin{table}[h!]
\centering
\begin{tabular}{ |c|c|c| } 
 \hline
 Substrates & Normalised Area & Raman Shift (cm$^{-1}$) \\ 
 \hline
 Pristine & 0.15 & 1252 \\ 
 Glass & 0.19 & 1247 \\
 PDMS & 0.21 & 1244 \\
 \hline
\end{tabular}
\caption{Characteristics of the Raman peak due to the inter-ring stretching mode of PEDOT:PSS in the pristine form and after patterning on different substrates}
\label{raman-1250}
\end{table}

Now, we consider the the wider peak in the vicinity of 1425 cm$^{-1}$. From the literature available, it appears that these peaks are due to a combination of two peaks, one around 1430 and the other at 1410 cm$^{-1}$, attributed to the benzoid and the quinoid conformation structures, respectively\cite{yang2011JAP}. The benzoid structures have been associated with a coil conformation of the polymer chain, whereas the quinoid structures have primarily been demonstrated to consist of linear conformations\cite{lenz2011ChemPh}. Since the charge conduction of the PEDOT:PSS sheet is better facilitated \cite{rutledge2015ACSAMI} with the linear conformation quinoid structure, that is the favoured configuration for high conductivity. In Fig.~\ref{Raman-cv}, the larger band around 1420 for all the samples have been deconvolved into two components with distinct modes of vibration, one at a higher wavenumber around 1420-30 cm$^{-1}$ and the other at a lower wavenumber of 1400-10 cm$^{-1}$. As has already been mentioned, the quinoid mode is represented by the lower wavenumber peak (red line) and the benzoid structure is represented by the higher wavenumber peak (green line). The analysis of the fitted quinoid and benzoid peaks as shown in Fig.~\ref{Raman-cv}b reveals that the broad Raman peak around 1425 cm$^{-1}$ for the pristine sample gets increasingly red shifted for glass and PDMS. Moreover, as we observe in Fig.~\ref{Raman-cv}b after deconvoluting the broad peak around 1420 cm$^{-1}$, the contribution of the quinoid peak gradually increases as we move from pristine to PDMS - with both peaks undergoing a red shift as well from 1433 to 1422 cm$^{-1}$. This indicates a structural rearrangement of the polymer chains, presumably as a result of micropatterning at different substrates, which causes a change in the conformational state of the PEDOT:PSS from largely being benzoid to quinoid structures. This is evident in the increasing ratio of the relative integrated intensities of the quinoid versus benzoid peaks  which we represent in Table~\ref{ratio}. Thus, pristine PSS-PEDOT has the lowest quinoid-benzoid ratio, while the samples patterned using MBL on PDMS have the highest. Given that a quinoid structure favours a linear conformation, PEDOT:PSS grains can be packed more densely, which increases their size and promotes more continuous grain boundaries \cite{funda2016JAP}. It is important to note that grain boundaries function as roadblocks that prevent the efficient transfer of charges. This is also proven by the SEM images where the grain boundaries become more and more homogeneous as we move from pristine to samples patterned on PDMS. Thus, due to the improved molecular order and dense packing, it is likely that interactions between polymers in the linear conformation will be greater than those between polymers in the coil conformation \cite{kim2016Elsevier}. Together, these factors will lead to the decrease of of the PEDOT:PSS sheet resistance from pristine to the patterned samples, with lowest resistance for the PDMS substrate.

\begin{table}[h!]
\centering
\begin{tabular}{ |c|c|c| } 
 \hline
  Substrates & Normalised Area & Raman Shift (cm$^{-1}$) of quinoid mode \\ 
 \hline
 Pristine & 0.8 & 1408 \\ 
 Glass & 2.5 & 1404 \\
 PDMS & 4.3 & 1402 \\
 \hline
 \end{tabular}
\caption{Characteristics of the Raman peak due to the quinoid mode in PEDOT:PSS in the pristine form and after patterning on different substrates.}
\label{ratio}
\end{table}

{\it Investigations to explain high conductivity: analysis of cyclic voltametry data:} The Raman spectroscopy data suggests that the PEDOT:PSS patterned on PDMS display a higher linear conformation of the constituent polymers due to the predominance of quinoid structures that support such linear conformations. This should also enhance charge delocalization on the PSS:PEDOT chains and raise carrier density. We analyse this issue in greater depths now, and attempt to understand the reason behind the linear conformation of PSS:PEDOT on PDMS substrates. For pristine PEDOT:PSS, we note that conductive PEDOT-rich cores are evenly buried in insulative PSS nanoshells. When exposed to MBL, this PEDOT core preferentially absorbs infrared photons and heats up significantly when irradiated with a 1064 nm laser beam. Following that, such PEDOT cores effectively transport this large amount heat energy to the PSS shell nearby. The heat released from the PEDOT core in the intense and immediate laser heating process then thermally fragments these PSS nanoshells, causing dynamic reorganisation. As a result, conductive PEDOT-rich cores emerge from locations where PSS shells previously existed, resulting in improved contact between surrounding conductive PEDOT-rich cores. This helps in reducing the carrier hopping distances between PEDOT rich cores and eventually leads to giant increase in the electrical conductivity \cite{yun2019generating}. Now, such restructuring also occurs at the patterning surfaces, which adds to the increase in conductivity. Since glass is a covalent interface, it is much more difficult to restructure than PDMS, which has a liquid polymeric interface that can be easily restructured during MBL. As a consequence of this restructuring,  more interfacial orientational order is generated, thus leading to higher conductivity of the PEDOT:PSS patterned on the PDMS substrate compared to that on the glass substrate. This also proves the importance of the substrate in improving the conductivity of patterned PEDOT:PSS. Note also, that if PSS is removed from the vicinity of PEDOT, the coulombic interactions between them would be reduced, causing PEDOT to shift conformation from coil to an extended-coil or a linear structure. Positive charges are then more delocalized as a result of these conformational changes in the PEDOT chain, which would also then contribute to conductivity improvements\cite{xia2012solution}.

In order to probe these effects, the electrochemical properties of the patterns need to be determined, which we carry out using cyclic voltametry. As evident from Fig.~\ref{Raman-cv}c the cyclic voltammograms of pristine PEDOT:PSS and patterns on glass and PDMS show oxidation peaks at 0.3 V, 0.13 V and -0.05V respectively. Similarly, reduction peaks are obtained at -0.7 V, 0 V and -0.3V, respectively. When compared to a pristine sample, the values of integrated current virtually double in the case of a pattern on glass, and rise by a factor of ten in the case of a pattern on PDMS. As a result, when compared to the other two, PDMS patterns produced the highest electrode current. This is clearly due to the fact that PDMS patterns have a higher amount of charge carriers than those on glass and the pristine form, which occurs due to the reasons mentioned earlier. 

The rectangular shape of the cyclic voltammograms of the patterns indicates a capacitive nature of the patterns. These redox features of the interface might correspond to the interconversion reactions happening between the oxidized and the reduced states of this conducting polymer due to patterning, while the shift in oxidation peaks can be explained in terms of interfacial ordering of the substrates arising upon laser irradiation during the patterning process. Since the 'liquid-like' interface of pure PEDOT:PSS is in a 'frozen' state in the interface, it allows for longer hopping distances. When patterned, however, the structural ordering in the matrices (both in glass and PDMS) squeezes the system into energy minima, resulting in shorter hopping times and easier electron transfers during interconversion reactions between the oxidised and reduced states, and also causing oxidation peaks to shift towards the lower potential region. Our hypothesis that the restructuring during MBL patterning on a PDMS matrix leads to more ordering than glass is validated by the observations that the lowest oxidation potential and highest conductivity is obtained in the case of the former.

\begin{figure}[!h]
\centering
\includegraphics[width=17cm]{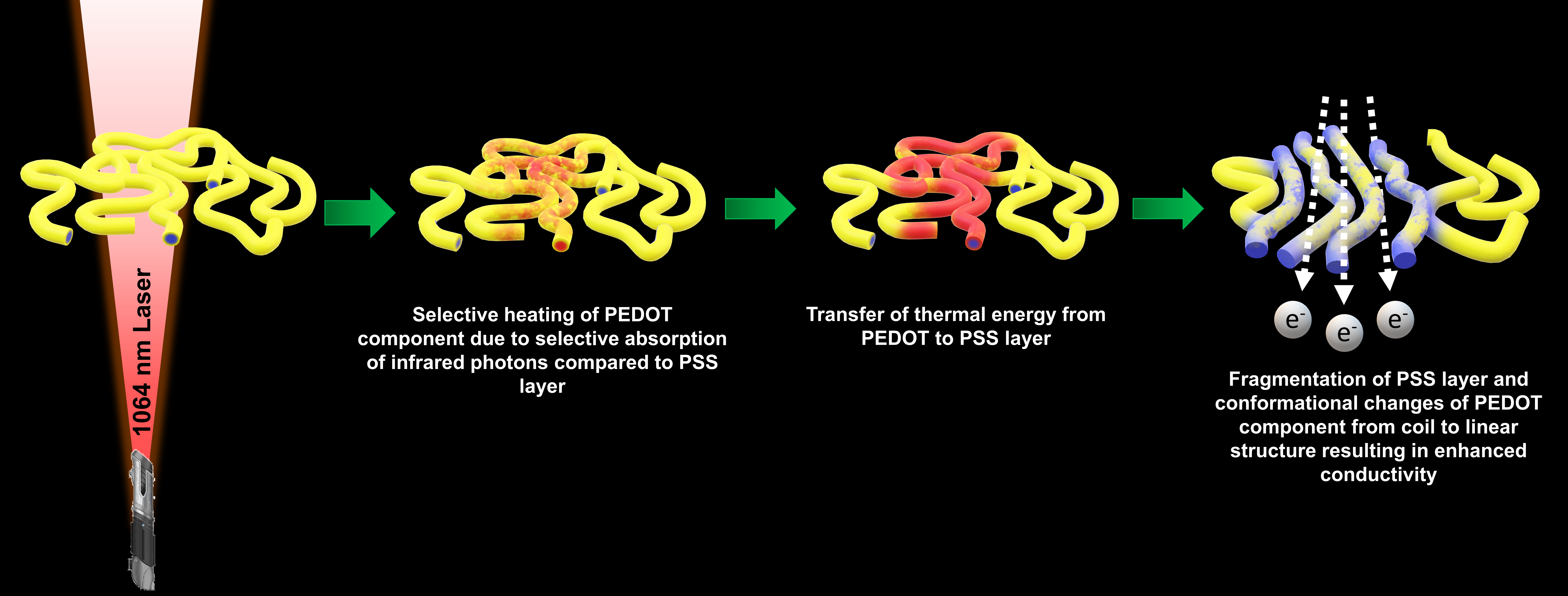}
\caption{Schematic representation of the mechanism of the conductivity enhancement for the PEDOT:PSS. PEDOT core selectively absorbs at the 1064 nm laser wavelength. This heat is eventually transferred to the nearest neighbour i.e PSS shells which results in their melting and the self assembly which causes this selective heating organises the polymers into linear coil structure.}
\label{conclude}
\end{figure}

{\it Conclusions}: We use MBL to carry out successful interface engineering of glass and PDMS substrates when we form stable linear patterns, so that we observe an enhancement of conductivity of around ten times for glass ($694\pm 24$ S/cm) and twenty times using PDMS ($2844\pm 321$ S/cm), compared to the pristine sample ($140\pm 12$ S/cm). Our method involves no external chemical treatment or doping of the substrates, and is intrinsic to the process of patterning itself. To determine the effect of the interfaces on the material patterned, and thus explain the enhanced conductivities,  we carry out a series of characterization experiments including SEM/EDX, Raman Spectroscopy, and cyclic voltametry. We conclude that MBL, due to the heat it generates, triggers a conformational change in the PEDOT:PSS polymers depending on the substrate on which they are patterned. Thus, the incident laser selectively heats up the PEDOT core due to the  strong selective absorption of the IR photons as compared to their PSS-affected surroundings as depicted in Fig.~\ref{conclude}. By heating up and melting the adjacent PSS shells as a result of this selective absorption, the PEDOT core becomes exposed to neighboring PEDOT cores. Along with the directional self-assembly, the fragmentation of PSS shells and lengthening of PEDOT cores leads to a coil (in pristine form) to an expanded coil (on glass) or even linear conformation (on PDMS) of the PEDOT:PSS, which improves conductivity. The interfacial ordering of the polymer is now further improved when this technique is applied to various surface matrices, which squeezes the system into energy minima and reduces hopping durations while facilitating quicker charge diffusion, all of which contribute to further enhancement of conductivity from pristine to glass to PDMS. 

This hypothesis is confirmed from the EDM images of the patterned PEDOT:PSS material, which display increasingly larger grain boundaries as we go from pristine to MBL patterns on glass and PDMS - with the boundaries virtually disappearing for the samples patterned on PDMS, thus allowing for larger charge diffusion - leading to higher conductivity for that case. Further confirmation is obtained from Raman spectroscopy, which probes deeper into the conformation of the polymer chains that make up PEDOT-PSS. We concentrate on two Raman modes - the inter-ring stretching mode at 1250 cm$^{-1}$, and the broad mode around 1425 cm$^{-1}$, which comprises of a convolution of benzoid and quinoid modes. We observe that the contribution of the quinoid mode - which is associated with linear polymer structures - increases as we go from pristine to glass and then PDMS substrate. Finally, we perform cyclic voltametry, where we observe that the value of electrode current is higher by a factor of two for PEDOT:PSS samples on glass, and almost ten for that on PDMS compared to the pristine form. This clearly demonstrates the availability of a higher number of charge carriers for PEDOT:PSS patterned on PDMS compared to that on the other interfaces. In addition, the oxidation peaks shift towards increasingly lower potentials for glass and PDMS, which can only occur due to structural ordering in both matrices, with higher ordering PDMS since it has a liquid polymeric interface that is easier to restructure in comparison to glass. This facilitates easier electron transfers due to shorter hopping times during interconversion reactions between oxidised and reduced states. Thus, the evidence we obtain from SEM and Raman spectroscopy that PDMS interfaces lead to more linear polymer structures are entirely validated by the data we obtain from cyclic voltametry. 
Indeed, the process of interface engineering that we drive by MBL to increase conductivity of PEDOT:PSS across glass and PDMS is summarized by the cartoon depicted in Fig.~\ref{conclude}.  

We believe that our experiments have provided a convincing argument to confirm that MBL leads to successful interface engineering which can change conductivity of conductive polymers rather drastically in the process of the patterning itself - which is further to the evidence we obtained earlier for this process simultaneously doping materials while patterning them to achieve similar conductivity increase for polypyrrole and polyaniline\cite{ghosh2017JMCC} micropatterns on glass. The next step is to attempt to build heterostructured interfaces which can lead to development of electronic devices such as diodes and transistors, which can take this extremely promising science for patterning micro or even nano-structures to the next level for value creation and subsequent translation to chip-fabrication technology. 

\begin{acknowledgement}
The authors thank IISER Kolkata, an autonomous research and teaching institution funded by the MHRD, Government of India for providing the financial support and infrastructure. The authors also thank Mithun Ajith and Gunaseelan M. of IIT Madras for useful discussions and help in the experiments. R.S. acknowledges DST for the INSPIRE fellowship.

\end{acknowledgement}
\bibliography{bibliography}
\end{document}